\pdfoutput=1

\documentclass[11pt]{article}

\usepackage{acl2023}

\usepackage{times}
\usepackage{latexsym}

\usepackage[T1]{fontenc}

\usepackage[utf8]{inputenc}

\usepackage{microtype}

\usepackage{inconsolata}
\usepackage{comment}
\usepackage{booktabs}

\usepackage{amsmath, amssymb, amsfonts, bbm}

\title{AlphaSharpe: LLM-Driven Discovery of Robust Risk-Adjusted Metrics}

\author{Kamer Ali Yuksel \and Hassan Sawaf \\
        aiXplain Inc., San Jose, CA, USA \\
  \texttt{\{kamer, hassan\}@aixplain.com} \\}
  
\begin{document}
\maketitle
\begin{abstract}
Financial metrics like the Sharpe ratio are pivotal in evaluating investment performance by balancing risk and return. However, traditional metrics often struggle with robustness and generalization, particularly in dynamic and volatile market conditions. This paper introduces AlphaSharpe, a novel framework leveraging large language models (LLMs) to iteratively evolve and optimize financial metrics to discover enhanced risk-return metrics that outperform traditional approaches in robustness and correlation with future performance metrics by employing iterative crossover, mutation, and evaluation. Key contributions of this work include: (1) a novel use of LLMs to generate and refine financial metrics with implicit domain-specific knowledge, (2) a scoring mechanism to ensure that evolved metrics generalize effectively to unseen data, and (3) an empirical demonstration of \textbf{3x} predictive power for future risk-returns, and \textbf{2x} portfolio performance. Experimental results in a real-world dataset highlight the superiority of discovered metrics, making them highly relevant to portfolio managers and financial decision-makers. This framework not only addresses the limitations of existing metrics but also showcases the potential of LLMs in advancing financial analytics, paving the way for informed and robust investment strategies.
\end{abstract}

\section{Introduction}

In finance, performance metrics such as the Sharpe ratio are pivotal in evaluating the trade-off between risk and return. The Sharpe ratio (the ratio of excess returns to the standard deviation of returns) has become a cornerstone of modern portfolio management due to its simplicity and widespread applicability. Yet, despite their popularity, the Sharpe ratio and similar metrics have several inherent limitations. There is a pressing need for financial performance metrics that are robust to market anomalies and capable of generalizing effectively to future scenarios. Such metrics should strongly correlate with future performance, providing actionable insights for portfolio managers and investors. Designing these advanced metrics, however, is a complex task that requires blending financial domain expertise with cutting-edge computational techniques.

Large Language Models (LLMs) present a unique opportunity to address these challenges. LLMs, trained on vast data corpora, exhibit an unparalleled ability to generate creative and contextually relevant solutions across domains. Their ability to analyze existing financial literature, synthesize insights, and propose novel approaches makes them ideal candidates for evolving financial metrics. Enhancing financial metric robustness has significant implications for portfolio management. Robust metrics enable more reliable decision-making by mitigating the impact of data noise and outliers. Furthermore, metrics that generalize to future scenarios help portfolio managers align their strategies with long-term performance goals. Integrating LLMs and iterative optimization represents a transformative step toward achieving these objectives.

This paper introduces \textbf{AlphaSharpe}, a novel framework for evolving robust financial metrics using LLMs. AlphaSharpe leverages the generative capabilities of LLMs to propose innovative financial metrics and employs iterative optimization techniques to refine and validate these metrics. The framework addresses traditional metrics' limitations by focusing on robustness, generalization, and predictive performance. The key contributions:

\begin{enumerate}
    \item A comprehensive system that utilizes LLMs for iterative refinement of investment performance metrics through crossover, mutation, and scoring to discover novel interpretable metrics with proven out-of-sample robustness.

    \item The innovative use of LLMs to generate diverse metric variants instilling financial literature and mathematical principles, enabling creativity in financial metric design.

    \item A structured methodology for evolving investment performance metrics through iterative mutation, crossover, and evaluation, ensuring their out-of-sample robustness validated by the alignment with future performance.

    \item Experiments demonstrating the robustness and predictive power of AlphaSharpe metrics compared to traditional metrics, highlighting their practical value in portfolio management.
\end{enumerate}

By addressing the limitations of traditional financial metrics and demonstrating the potential of LLMs in this domain, AlphaSharpe represents a significant advancement in financial analytics. This work lays the foundation for further exploration into integrating LLM-driven methodologies to enhance financial decision-making.

\section{Background}

Financial metrics are critical for evaluating investment performance \citep{markowitz}. Metrics like the \textit{Sharpe Ratio} by \citet{sharpe1966mutual}, have long been used to assess risk-adjusted returns but face notable limitations. While widely adopted, they are sensitive to outliers, rely on normality assumptions, and lack robustness in non-stationary environments.

\begin{itemize} \item \textbf{Sensitivity to Outliers:} Metrics like the Sharpe ratio can be skewed by extreme values in return distributions \citep{bailey2014deflated}. \item \textbf{Stationarity Assumptions:} These metrics assume a static risk-return relationship, which may not hold in dynamic markets \citep{bailey2012sharpe}. \item \textbf{Backward-Looking Nature:} They evaluate past performance without adequately correlating with future outcomes. \item \textbf{Limited Generalization:} Metrics often fail to adapt to diverse asset classes or market conditions, leading to suboptimal decisions. \end{itemize}

An improvement over the traditional Sharpe ratio, the Probabilistic Sharpe Ratio (PSR), incorporates statistical inference to account for uncertainty in performance evaluation \citep{bailey2014deflated}. By considering the distribution of Sharpe ratio estimates, the PSR reduces noise and offers a probabilistic interpretation of performance consistency. Human-discovered innovations in financial investment performance metrics like the PSR are revolutionary but still suboptimal in robustness and predictive generalization.

Machine learning (ML) has transformed financial analysis by introducing adaptive and predictive capabilities that traditional approaches lack. It has been widely applied in portfolio optimization, risk management, and metric design. ML models, like Deep Neural Networks (DNNs), and reinforcement learning enable learning dynamic asset allocation strategies. These models optimize for risk-adjusted returns under varying market conditions, outperforming static portfolio strategies. DNNs can be used to train nonlinear metrics that capture complex relationships in return data, offering enhanced predictive accuracy over traditional methods. For example, DNNs trained on historical data optimize for metrics that maximize future performance consistency. Yet, these approaches would be heavily prone to overfitting and lack interpretability.

Recent advancements in reinforcement learning, particularly through AlphaTensor and AlphaCode, provide a blueprint for iterative optimization and creative problem-solving. AlphaTensor, introduced by \citet{fawzi2022discovering}, uses multi-agent reinforcement learning to optimize algorithms iteratively, discovering efficient methods for matrix multiplication. Its iterative optimization framework inspired the proposed approach for evolving financial metrics. Similarly, AlphaSharpe employs LLMs to propose, score, and refine financial metrics through an iterative workflow, ensuring continuous improvement. AlphaCode, demonstrated by \citet{li2022competition}, highlights the potential of large-scale models in generating high-quality code for competitive programming tasks. Its ability to learn from examples and refine outputs through iterative feedback strongly parallels the generation of financial metrics optimized for out-of-sample robustness.

Recent works such as \citet{romera2024mathematical}, also demonstrated how LLMs can contribute to mathematical discovery by automating the search for novel programs and equations, and \citet{lu2024ai}, who proposed a framework for fully automated, open-ended scientific discovery using LLMs. These works further highlighted the growing potential of AI to drive innovation in scientific and technical domains, reinforcing the potential of LLM-driven iterative workflows like AlphaSharpe to redefine financial metric design.

\section{Methodology}

LLMs, such as OpenAI’s GPT, Meta’s LLama, have demonstrated exceptional capabilities in generating codes for creative and robust algorithms across several domains. In this work, LLMs are used to revolutionize financial metric discovery by:
\begin{itemize}
    \item \textbf{Few-Shot Generation:} Leveraging few-shot examples to generate creative variations of existing and discovered metrics, combining domain expertise with data-driven insights.
    \item \textbf{Iterative Optimization:} Utilizing evolutionary strategies and feedback loops to iteratively refine generated metrics for better robustness and predictive power.
    \item \textbf{Cross-Domain Inspiration:} Drawing on extensive training data across domains, LLMs can introduce concepts from other disciplines to innovate financial metrics.
    \item \textbf{Mutational Refinement:} Inspired by evolutionary algorithms, LLMs can suggest mutations to existing metrics, optimizing robustness and generalization.
    \item \textbf{Automated Code Generation:} Similar to AlphaCode, LLMs can generate high-quality implementations for new financial metrics.
    \item \textbf{Critical Thinking and Synthesis:} By analyzing the academic literature, LLMs can integrate theoretical insights into metric design, ensuring both novelty and rigor.
\end{itemize}

The AlphaSharpe is a novel method designed to iteratively optimize financial performance metrics, such as the Sharpe ratio, for out-of-sample robustness by leveraging large language models (LLMs) for creativity and critical thinking. The framework employs an evolutionary approach that blends the implicit domain expertise of LLMs and evolutionary strategies to design metrics that exhibit superior robustness and generalization capabilities. LLMs, such as GPT-based models, generate novel metrics by drawing inspiration from academic literature and best practices in financial analysis. Few-shot learning and prompt engineering guide LLMs in producing relevant and innovative metrics.

\subsection{Architecture}
The workflow comprises an iterative four-step process that refines financial metrics through crossover, mutation, scoring, and ranking (selection). Each iteration is designed to improve the metrics' robustness and predictive capabilities incrementally. The framework operates in a looped pipeline, where each iteration refines existing metrics or generates new ones through crossover and mutation:
\begin{itemize}
    \item \textit{Crossover:} Combining elements from a diverse set of top-performing metrics to create hybrids that inherit strengths from all of them.
    \item \textit{Mutation:} LLM generates meaningful variations of crossovered metric by making small but deliberate modifications to enhance predictive capability (out-of-sample robustness).
\end{itemize}

Mutated metrics are evaluated using scoring functions based on:
        \begin{itemize}
            \item \textit{Robustness:} Sensitivity to outliers and extreme market conditions.
            \item \textit{Generalization:} Correlation between the metric scores on historical data and future performance (e.g., future Sharpe ratios).
            \item \textit{Predictive Power:} Statistical relevance in identifying high-performing assets.
        \end{itemize}

Metric quality is evaluated using statistical criteria such as robustness, correlation with future Sharpe ratios, and normalized discounted cumulative gain (NDCG). These functions ensure that the designed metrics generalize effectively to unseen scenarios. Metrics are ranked based on quality and diversity; only top candidates are retained for cross-over in further iterations. Cross-over combines these top-ranked metrics from the scoring phase to create hybrids blending their computational elements. This process leverages the strengths of individual metrics while mitigating their weaknesses. Metrics that rank poorly are discarded, ensuring only a diverse set of high-quality candidates proceed to further stages \citep{cully2017quality}.

\subsection{LLM Usage}
LLMs are integral to the AlphaSharpe, serving as the creative engine for metric generation and refinement. These models enable the system to draw from vast repositories of financial knowledge and provide innovative solutions that traditional methods might overlook. They are trained to understand the structure and purpose of financial metrics based on context from research papers, textbooks, and industry practices. They simulate the thought processes of domain experts, proposing metrics inspired by existing methodologies while introducing novel elements to discover unexplored dimensions. AlphaSharpe balances domain expertise and computational innovation by integrating LLMs into the workflow, making it a powerful tool for evolving robust financial metrics. Together, these components enable it to evolve metrics iteratively, pushing the boundaries of investment analysis.

To effectively harness the potential of LLMs, the workflow employs carefully designed prompts that: (1) guide the LLM to integrate domain-specific insights with contextual inputs, such as risk management principles or portfolio strategies (2) encourage innovation and critical-thinking by asking the LLM to "think outside the box" and propose metrics generalizing better to future (3) emphasize using tensor operations and avoiding resource-intensive loops or redundant hyperparameters.

\subsection{Scoring Functions}
AlphaSharpe employs robust scoring mechanisms to evaluate and evolve financial metrics. These functions measure how well a proposed metric generalizes to the future (out-of-sample) performance. To evaluate and rank top-performing metrics, AlphaSharpe employs the following workflow: (1) Apply the proposed metric to a historical asset log-return dataset to compute scores. (2) Evaluate the alignment of scores with future Sharpe ratios via:

\begin{itemize} \item \textbf{Spearman’s Rho}: Measures the monotonic relationship between rankings produced by the metrics on historical data and the realized future Sharpe ratios \citep{spearman1904proof}. \item \textbf{Kendall’s Tau}: Assesses the strength of ordinal associations between metric scores and future Sharpe ratios, offering a robust ranking correlation metric \citep{kendall1938new}. \item \textbf{Normalized Discounted Cumulative Gain (NDCG)}: Evaluates the quality of asset rankings, placing greater importance on correctly ranking top-performing assets, critical in financial contexts such as portfolio selection \citep{jarvelin2002cumulated}. \end{itemize}

These metrics evaluate how consistently the rankings of assets based on the metric align with their rankings based on future performance. They reduce sensitivity to outliers and non-linearities by relying on the ranks of data instead of raw values. 

\section{Experiments}

Time-series cross-validation was employed to validate and rank metrics as they evolve. The dataset, comprising 15 years of historical data from 3,246 US stocks and ETFs, was split into overlapping folds after separating 20\% of the data for the out-of-sample test (3 years). Metrics evolved based on their correlation with future Sharpe ratios within the cross-validation sets, ensuring robust evaluation across different periods. The evolved metrics were finally blind-tested during recent periods of extreme market stress, including the 2020 COVID-19 market crash, highlighting their robustness and stability under high-volatility conditions \citep{lipton2020three}. Evolved to correlate better with future Sharpe ratios by iteratively refining their formulation through LLM-driven mutation and scoring. The resulting metric adapts to varying market conditions and improves predictive accuracy for future Sharpe ratios, as resulted in experiments. The implementation of the discovered metrics, \textbf{AlphaSharpe Ratio} (\( \alpha_S \)), and the code for reproducing the experiments are \href{https://anonymous.4open.science/r/alphasharpe}{open-sourced}.

The $\alpha_{S1}$ to $\alpha_{S4}$ metrics are advanced alternatives to the traditional Sharpe ratio, addressing its limitations in handling small sample sizes, extreme values, and non-normal return distributions. These metrics provide a robust and realistic assessment of risk-adjusted performance, ensuring greater reliability for evaluating modern financial strategies.

\[
\alpha_{S1} = \frac{\exp\left(\mathbb{E}[\log R - r_f]\right)}{\sqrt{(\sigma_{\log R}^2 + \epsilon) \cdot (\sigma_{\log R} + \epsilon)}}
\]
$\alpha_{S1}$ uses the expected log excess return to emphasize compounding effects, while incorporating stability constants ($\epsilon$) to reduce sensitivity to extreme values and ensure robustness in small datasets.

\[
\alpha_{S2} = \frac{\exp\left(\mathbb{E}[\log R - r_f]\right)}{\sqrt{\sigma_{\log R}^2 + \epsilon} + DR + V}
\]

\textbf{Downside Risk Adjustment:} The $\alpha_{S2}$ metric builds on its parent $\alpha_{S1}$ by integrating downside risk. This addition accounts for negative returns, penalizing them more heavily and better reflecting real-world risk asymmetries.

\[
DR = \frac{\sigma_{R^-} + \sqrt{N_{R^-}} \cdot \sigma_{\log R}}{N_{R^-} + \epsilon}
\]

\textbf{Forecasted Volatility Adjustment:} \( \alpha_{S2} \) also introduces forecasted volatility (\(\text{V}\)) as an additional component for forward-looking risk assessment:
\[
\text{V} = \sqrt{\frac{1}{n} \sum_{t=n/4}^{n} (R_t - \mathbb{E}[R])^2}
\]

\textbf{Higher-Order Moment Adjustment:} The \( \alpha_{S3} \) metric further evolves from the \( \alpha_{S2} \) metric by incorporating higher-order moments of return distributions ($HAD$), such as skewness and kurtosis, along with a dynamic adjustment for drawdown risks. This enables \( \alpha_{S3} \) to more comprehensively evaluate risk-adjusted performance.

\[
\alpha_{S3} = \alpha_{S2} \cdot \left(1 - \frac{K}{12}\right) \cdot \frac{1 + \frac{S}{6}}{1 + MDD}
\]

\textbf{Regime Factor Adjustment:} \( \alpha_{S4} \) extends the evaluation by incorporating a regime-dependent factor to dynamically adjust for market conditions:
\[
\alpha_{S4} = \alpha_{S3} \cdot \left(1 + 0.1 \cdot \mathbb{1}(\mathbb{E}[R - r_f] > 0)\right)
\]

These metrics, \( \alpha_{S1} \) through \( \alpha_{S4} \), are particularly suitable for evaluating strategies with log-transformed returns, especially in volatile markets or scenarios with skewed and fat-tailed return distributions. By addressing key limitations of the traditional Sharpe ratio, these advanced metrics offer nuanced, reliable perspectives for assessing risk-adjusted performance. The experimental results demonstrate the significant advantages of AlphaSharpe metrics over traditional financial metrics in ranking correlation and portfolio construction.

\subsection{Ranking Correlations}

AlphaSharpe achieved significantly higher ranking correlations than traditional metrics. Table~\ref{tab:ranking_correlation} shows the Spearman correlation, Kendall correlation, and NDCG@25\% scores for discovered metrics where the empirical results showcase AlphaSharpe's significant enhancement in ranking correlations.

\begin{table}[h]
\centering
\caption{Asset Ranking Correlation Comparison}
\label{tab:ranking_correlation}
\begin{tabular}{lccc}
\toprule
\textbf{Metric} & \textbf{Spearman} & \textbf{Kendall} & \textbf{NDCG} \\
\midrule
Sharpe Ratio    & 0.130 & 0.087 & 0.393 \\
PSR             & 0.127 & 0.085 & 0.388 \\
\( \alpha_{S1} \)    & 0.390 & 0.265 & 0.594 \\
\( \alpha_{S2} \)    & 0.394 & 0.268 & \textbf{0.636} \\
\( \alpha_{S3} \)    & 0.400 & 0.272 & 0.590 \\
\( \alpha_{S4} \)    & \textbf{0.409} & \textbf{0.278} & 0.589 \\
\bottomrule
\end{tabular}
\end{table}

\begin{itemize}
    \item \textbf{Exceptional Ranking Quality:} AlphaSharpe achieved over 3x higher Spearman and Kendall correlations than traditional metrics, providing unparalleled asset ranking accuracy.
    \item \textbf{Spearman's Rho:} \( \alpha_{S4} \) achieves a correlation of 0.409, which is \textbf{over 3x higher} than the Sharpe Ratio and PSR, demonstrating its robustness in ranking assets effectively.
    \item \textbf{Kendall's Tau:} \( \alpha_{S4} \) records a correlation of 0.278, maintaining \textbf{over 3x improvement} compared to traditional metrics, further emphasizing its reliability in asset ranking.
    \item \textbf{NDCG@25\% Score:} \( \alpha_{S2} \) delivers a substantial improvement, achieving a score of 0.636, reflecting a \textbf{+62\% increase} over the Sharpe Ratio and PSR, underscoring its superior ability to prioritize top-performing assets.
\end{itemize}

\noindent These empirical results on out-of-sample tests highlight AlphaSharpe metrics's exceptional capability to rank assets more accurately and reliably, aligning rankings closely with their future performance and demonstrating their value in financial analysis.

\begin{table}[h]
\centering
\caption{Portfolio Performance Improvement of \( \alpha_{S2} \) }
\label{tab:portfolio_performance}
\begin{tabular}{lcc}
\toprule
Threshold (\%) & $\Delta_{\text{Sharpe}}$ (\%) & $\Delta_{\text{PSR}}$ (\%) \\
\midrule
10\% & +76.49 & +82.17 \\
15\% & +91.47 & +93.69 \\
20\% & +93.39 & +95.61 \\
25\% & \textbf{+93.97} & \textbf{+101.99} \\
\bottomrule
\end{tabular}
\end{table}

\subsection{Portfolio Construction}

To assess the practical utility of AlphaSharpe, portfolios were constructed by selecting the top-performing assets ranked by each metric. Portfolios were created for the top 10\%, 15\%, 20\%, and 25\% of assets. Uniform weights were assigned to all selected assets in the portfolio, ensuring equal exposure to each asset \citep{demiguel2009optimal}. The test performance of portfolios was then evaluated using the Sharpe Ratio.  Table~\ref{tab:portfolio_performance} summarizes the percentage improvements of \( \alpha_S \) over the Sharpe Ratio and PSR regarding the resulting Sharpe Ratio on the test period. AlphaSharpe achieved a +89.11\% improvement over the Sharpe Ratio and a +95.73\% improvement over PSR, consistently delivering significant improvements across thresholds.

The results demonstrate that AlphaSharpe metrics significantly enhance financial evaluations' robustness, predictive accuracy, and practical utility. These evolved metrics consistently outperformed traditional measures in ranking correlations and portfolio performance, providing a groundbreaking tool for portfolio managers and financial analysts.

\begin{itemize}
    \item \textbf{Superior Portfolio Performance:} Portfolios constructed using \( \alpha_{S2} \) consistently achieved significantly higher Sharpe Ratios, with improvements of up to \textbf{+93.97\%} and \textbf{+101.99\%}, enhancing risk-adjusted returns.
    \item \textbf{Consistency Across Thresholds:} \( \alpha_{S2} \)'s robust performance was observed across varying portfolio sizes, from the top 10\% to 25\% of assets, making it a highly adaptable and scalable tool for diverse investment strategies.
\end{itemize}

\section{Implications}

AlphaSharpe metrics redefine the evaluation of financial strategies by addressing the limitations of traditional measures like the Sharpe Ratio. Their ability to handle small datasets, integrate distributional nuances (such as skewness and kurtosis), and adapt to changing market regimes makes them invaluable for modern portfolio management and risk assessment. The metrics' empirical outperformance demonstrates their reliability in ranking assets and constructing portfolios with superior risk-adjusted returns. As financial markets grow increasingly complex, tools like AlphaSharpe offer a forward-looking, data-driven approach to optimizing investment strategies, ensuring resilience and adaptability in an ever-changing landscape.

\begin{itemize}
    \item \textbf{Enhanced Portfolio Construction:}  
    The advanced adjustments within AlphaSharpe metrics enable more precise selection and allocation of portfolio assets, particularly in scenarios involving volatile markets or non-normal return distributions. This ensures robust performance across diverse investment strategies.

    \item \textbf{Improved Risk Management:}  
    AlphaSharpe metrics provide a more realistic assessment of risk-adjusted performance by incorporating downside risks, higher-order moments, and dynamic adjustments for market regimes. Investors can better prepare for extreme market events and mitigate potential losses.

    \item \textbf{Dynamic Adaptability:}  
    The regime-dependent factors in \( \alpha_{S4} \) allow for dynamic portfolio adjustments, making it ideal for strategies that respond proactively to shifting market conditions and economic cycles.

    \item \textbf{Broader Applicability:}  
    AlphaSharpe metrics’ ability to outperform traditional measures across ranking, risk assessment, and portfolio optimization makes them a versatile tool for financial analysts, portfolio managers, and quantitative researchers.
\end{itemize}

\section{Discussion}

The LLM-driven evolution blended elements from traditional metrics to address the Sharpe Ratio's limitations. These metrics enhance robustness, generalization, and predictive power by incorporating innovations such as log-transformed returns, downside risk adjustments, higher-order moments, and regime-sensitive factors, making them relevant in financial analysis. \( \alpha_{S1} \) addresses the Sharpe Ratio's sensitivity to outliers and normality assumptions by using log returns and stability constants. Its focus on compounding effects also aligns with the Kelly Criterion \citep{kelly1956new} principles. \( \alpha_{S2} \) further extends this foundation by introducing downside risk adjustments, drawing inspiration from the Sortino Ratio \citep{sortino1994performance} and tail-risk metrics like Value-at-Risk (VaR) and Conditional VAR \citep{artzner1999coherent}.

\( \alpha_{S3} \) incorporates skewness, kurtosis, and maximum drawdown (\( MDD \)) to account for higher-order moments and extreme losses, inspired by metrics such as the Skewness-Adjusted Sharpe Ratio \citep{zakamulin2015beyond}, Calmar Ratio, and Sterling Ratio \citep{young1991calmar}. These additions enable \( \alpha_{S3} \) to handle non-normal return distributions and periods of high volatility, similar to the Omega Ratio’s focus on capturing the full return distribution \citep{keating2002universal}. Finally, \( \alpha_{S4} \) incorporates regime-dependent adjustments, dynamically adapting to market conditions such as positive or negative excess returns. This aligns with regime-switching models like Markov frameworks \citep{hamilton1989new} and extends the adaptability of risk-return metrics with state-dependent factors.

\section{Portfolio Optimization}

Traditional portfolio allocation techniques like the Risk Parity Portfolio and Equal Risk Contribution Portfolio often struggle with stability, extreme return sensitivity, and risk-adjusted return maximization inefficiencies \cite{qian2005risk, maillard2010properties}. The LLM-driven discovery of portfolio allocation methods has led to developing a novel approach—\textbf{AlphaSharpe Portfolio}. This newly discovered method significantly enhances portfolio performance by integrating inverse covariance risk-adjusted returns, stability-adjusted weighting, entropy regularization, and volatility normalization. The comparison in Table~\ref{tab:performance_comparison} demonstrates that it achieves the highest Sharpe Ratio and Calmar Ratio, outperforming traditional portfolio optimization techniques by a significant margin. The code for reproducing the experiments is \href{https://anonymous.4open.science/r/alphasharpe}{open-sourced}.

\section*{AlphaSharpe Portfolio}

The newly discovered AlphaSharpe Portfolio optimization method follows a structured approach, where it allocates a long-only portfolio by adjusting weights for stability, volatility, and diversification while preserving an optimal risk-return profile.

\begin{enumerate}
    \item Compute the \textbf{mean excess return vector} and the \textbf{covariance matrix}, and the \textbf{inverse covariance-adjusted risk-return vector}. \( R \) is the excess log return matrix, \( \mu \) is the mean excess return vector, and \( \lambda \) is a constant.
    \[
    \mu = \frac{1}{T} \sum_{t=1}^{T} R_t, \quad \Sigma = \text{Cov}(R) + \lambda I
    \]
    \[
    r = \max(0, \Sigma^{-1} \mu)
    \]
    \item Introduce a \textbf{stability factor} by incorporating the standard deviation of the inverse covariance-adjusted risk-return vector, apply \textbf{volatility normalization} and \textbf{softmax normalization} to derive initial asset allocations.
    \[
    r' = \frac{1 + \text{std}(r) \cdot r}{\sqrt{\text{diag}(\Sigma) + \epsilon}}
    \]

    \item Introduce \textbf{entropy regularization} to ensure diversification penalizing dominant assets:
    \[
    H = -\sum_i w_i \log(w_i + \epsilon)
    \]
    \[
    w' = \frac{\exp(r')}{\sum_i \exp(r'_i)} \cdot e^{-H}
    \]

    \item Compute \textbf{final portfolio weights}, ensuring risk-balanced allocation while preventing excessive concentration:
    \[
    w^* = \frac{\max(0, w')}{\sum_i \max(0, w')}
    \]
\end{enumerate}

The effectiveness of AlphaSharpe Portfolio is evident in its superior performance across key financial metrics compared to conventional strategies. Table~\ref{tab:performance_comparison} presents the Sharpe and Calmar Ratio \cite{young1991calmar} improvements over the Equal-Weighted Portfolio \citep{demiguel2009optimal}.

\begin{table}
\centering
\caption{Performance improvement of different portfolio strategies relative to the Equal Weighted benchmark.}
\label{tab:performance_comparison}
\begin{tabular}{lcc}
\toprule
\textbf{Portfolio Strategy} & $\Delta_{\text{Sharpe}}$ (\%) & $\Delta_{\text{Calmar}}$ (\%) \\
\midrule
Equal Weighted & 0.00\% & 0.00\% \\
Risk Parity & +38.32\% & +10.36\% \\
Equal Risk Cont. & +38.55\% & +10.44\% \\
AlphaSharpe & \textbf{+71.04}\% & \textbf{+116.31}\% \\
\bottomrule
\end{tabular}
\caption{Performance improvement of different portfolio strategies relative to the Equal Weighted benchmark.}
\label{tab:performance_comparison}
\end{table}

\begin{itemize}
\item \textbf{Highest Sharpe Ratio}: Achieves a 71.04\% increase over the Equal Weighted Portfolio, demonstrating superior risk-adjusted returns.
\item \textbf{Highest Calmar Ratio}: Records a 116.31\% improvement, signifying a more resilient strategy with lower drawdowns.
\item \textbf{Enhanced Stability}: The method’s stability factor prevents excessive concentration in high-risk assets, ensuring robustness.
\item \textbf{Entropy-Based Diversification}: Encourages a balanced risk contribution across assets, reducing reliance on volatile securities.
\end{itemize}

The AlphaSharpe Portfolio outperforms traditional portfolio strategies in risk-adjusted and drawdown-adjusted performance. This showcases the power of LLM-driven discovery of a portfolio optimization method, redefining the landscape through adaptive, stability-enhanced, and entropy-regularized portfolio optimization techniques.

\section{Conclusion}
This paper introduced AlphaSharpe, a novel framework that leverages LLMs to iteratively evolve robust financial metrics. AlphaSharpe addresses key limitations of traditional risk-adjusted metrics, including limited robustness, poor generalization, and insufficient predictive accuracy; by integrating LLMs for metric design, refinement, and optimization. The evolution of financial metrics demonstrated the power of combining LLM-driven creativity with a structured optimization framework. LLMs played a pivotal role by generating innovative, interpretable metrics inspired by financial literature and mathematical principles. Their ability to process few-shot examples and extrapolate ideas enabled the discovery of solutions that balance novelty with practical relevance, complementing human expertise and accelerating innovation.

AlphaSharpe metrics significantly improve ranking accuracy and portfolio performance, with adjustments that address volatility, skewness, and market regime shifts. For financial institutions, adopting these metrics enhances decision-making, fosters trust, and delivers competitive advantages by enabling more precise asset selection and allocation, even in dynamic or volatile markets. In conclusion, AlphaSharpe represents a transformative advancement in financial analytics, showcasing the potential of LLMs to redefine risk-return evaluation. By addressing the limitations of traditional metrics and introducing adaptive, robust alternatives, AlphaSharpe paves the way for more informed and resilient investment strategies. Future work can focus on enhancing interpretability and scaling its application across broader use cases.

\bibliography{custom}
\bibliographystyle{acl_natbib}

\end{document}